\documentclass{article}
\usepackage{amsmath,amssymb,enumerate,eucal,theorem}

\numberwithin{equation}{section}

\theoremheaderfont{\normalfont\scshape}
{\theorembodyfont{\rmfamily}}
{\theorembodyfont{\rmfamily}}
\newtheorem{conj}{Conjecture}[section]
\newtheorem{prop}{Proposition}[section]
\newtheorem{thm}{Theorem}[section]

\DeclareMathOperator{\ch}{ch}
\DeclareMathOperator{\disc}{disc}
\DeclareMathOperator{\rank}{rank}
\DeclareMathOperator{\Sym}{Sym}
\DeclareMathOperator{\Tr}{Tr}

\def\mn{\medskip\noindent}
\def\pn{\par\noindent}
\def\sn{\smallskip\noindent}

\title{Hecke-Langlands Duality and Witten's Gravitational Moonshine}
\author{Igor Yu. Potemine}
\date{}

\begin{document}
\maketitle
\abstract{We show that there is a dual description of conformal blocks of $d=2$ rational CFT in terms of Hecke eigenfields and eigensheaves. In particular, partition functions, conformal characters and lattice theta functions may be reconstructed from the action of Hecke operators. This method can be applied to: 1) rings of integers of Galois number fields equipped with the trace (or anti-trace) form; 2) root lattices of affine Kac-Moody algebras and WZW-models; 3) minimal models of Belavin-Polyakov-Zamolodchikov and related $d=2$ spin-chain/lattice models; 4) vertex algebras of Leech and Niemeier lattices
and others. We also use the original Witten's idea to construct the
3-dimensional quantum gravity as the AdS/CFT-dual of $c=24$ Monster vertex algebra of Frenkel-Lepowsky-Meurman. Concerning the geometric Langlands duality, we use results of Beilinson-Drinfeld, Frenkel-Ben-Zvi, Gukov-Kapustin-Witten and many others (cf. references).}

\tableofcontents

\section{Introduction. Number-theoretical dualities, $S$-dualities and gravitational cosmology}

In this paper we discuss an amazing interplay between number-theoretical dualites and black holes in Anti-de Sitter spaces. It passes through modular invariance of conformal field theories and AdS/CFT correspondences.

\smallskip Recall that electric $E$ and magnetic field $B$ in the vacuum (in regions without charges and currents) are related by Maxwell equations, invariant by transformations:
\begin{equation}
E\mapsto B,\ B\mapsto (-1/c^2)E
\end{equation}
where $c$ is the speed of light. This is the basic form of the electromagnetic duality. We can already interpret it as the modular 
$S$-invariance corresponding to the matrix $S=\begin{pmatrix}0 &-1\\ 1/c^2 &0\end{pmatrix}$, exchanging fields and reversing coupling constants. It extends to $S$-dualities in CFT and string theories.

\smallskip In recent years \cite{GW,KW}, it was realized that the Langlands correspondence is the number-theoretical counterpart of the electromagnetic duality in gauge field theories. Representations of the absolute Galois group correspond to modular objects on the moduli space of curves. Moreover, these modular objects are eigensheaves of
Hecke operators with meaningful eigenvalues. In particular, the coefficients of the $L$-function
\begin{equation}
L(E,s) = \sum_{n=1}^{\infty} \frac{a_n}{n^s}
\end{equation}
of an elliptic curve $E/\mathbb{Q}$ are Hecke eigenvalues, generating a cusp modular form 
\begin{equation}
f(E,q) = \sum_{n=1}^{\infty} a_n q^n
\end{equation}
of weight 2. This is a consequence of the famous \emph{modularity theorem} related to the Fermat's Last Theorem.

\smallskip In this paper we define vertex operator algebras for all Galois number fields $K/\mathbb{Q}$, equipped with integral trace forms on their rings of integers ${\mathcal O}_K$ (\emph{cf.}~sect.~5). Thus, the modular $S$-invariance of lattice CFTs is closely related to the arithmetic Langlands program.

\smallskip In addition, Witten \cite{Wit} has associated, via AdS/CFT correspondence, the pure $d=3$ gravity to the \emph{Monster Moonshine Module} $V^\natural$. It transports the \emph{Hecke-Langlands duality} for the extremal $c=24$ CFT to modular dualities for BTZ black holes on 
AdS$_3$ (\emph{cf.}~sect.~11).

\section{Virasoro algebra and rational CFT}

For general mathematical definitions, related to vertex algebras and conformal field theories (CFT), see \cite{FBZ, Kac}.

\par Recall that a vertex algebra $V=\oplus V_n$ is conformal of central charge
$c\in\mathbb{C}$ if it contains a conformal vector $\omega\in V_2$ such that Fourier coefficients $L_n=L_n^V$ of the corresponding vertex operator
\begin{equation}
Y(\omega,z) = \sum L_n^V z^{-n-2}
\end{equation}
satisfy the defining relations of the Virasoro algebra:
\begin{equation}
[L_n,L_m] = (n-m)L_{n+m} + c\,\frac{n^3-n}{12}\,\delta_{n,-m},
[L_n,c]=0.
\end{equation}
In addition, $L_{-1}^V=T:V\rightarrow V$ should be a translation operator of degree 1, $L_0^V|_{V_n} = n\cdot\mathrm{Id}$ the grading operator and $L_2\omega = \frac{c}{2}|0\rangle$.

\smallskip The Virasoro algebra is the central extension of 
$\mathbb{C}((t))\partial_t$ with topological basis given by 
$L_n=-t^{n+1}\partial_t$, $n\in\mathbb{Z}$, and $c\in\mathbb{C}$.

\smallskip A conformal vertex algebra V is called \emph{rational} if it is completely reducible with finitely many inequivalent simple modules having finite-dimensional graded components. 

\smallskip The basic example is given by the \emph{Virasoro vertex algebra} $\mathrm{Vir}_c$ with central charge $c$ and conformal vector $\omega = L_{-2}|0\rangle$ \cite[sect.\,2.5]{FBZ}.

\section{Lattice vertex superalgebras}

Let $\Lambda$ be a lattice in $\mathbb{R}^n$ equipped with a non-degenerate $\mathbb{Z}$-valued symmetric bilinear form $\langle
\cdot,\cdot\rangle$. There is a well-known quantization of the space of maps from the circle $S^1$ to the torus $\mathbb{R}^n/\Lambda$ called \emph{lattice vertex superalgebra}.

\smallskip Denote by $\mathfrak{h}=\mathbb{C}\otimes_{\mathbb{Z}} \Lambda$ the complexification of $\Lambda$ and $\mathfrak{h}^{<0}=\sum_{j<0} h\otimes t^j$. The space of state is
\begin{equation}
V_{\Lambda} = \Sym\left(\mathfrak{h}^{<0}\right)\otimes C[\Lambda]
\end{equation}
where $\mathbb{C}[\Lambda]$ is the group algebra of $\Lambda$, generated by $e^\alpha$ ($\alpha\in\Lambda$). It is a superspace with parity $p(e^\alpha)=p(\alpha)=\langle\alpha,\alpha\rangle \mod 2$.

\smallskip For any $\alpha\in L$, mutually local fields $\Gamma_\alpha (z) =Y(1\otimes e^\alpha)$ are of the form
\begin{equation}
\Gamma_\alpha (z)= e^\alpha\exp\left({-\sum_{j<0}\frac{z^{-j}}{j} \alpha_j}\right) \exp\left({-\sum_{j>0}\frac{z^{-j}}{j} \alpha_j}\right) c_\alpha z^{\alpha_0}
\end{equation}
(\emph{cf.}~\cite[sect.~5.4]{Kac}) with operators $c_\alpha$ on $V_L$ satisfying
\begin{equation}
c_0=1,\ c_\alpha|0\rangle=|0\rangle,\ [h_n, c_\alpha]=0\ 
(h\in\mathfrak{h},\ n\in\mathbb{Z})
\end{equation}
\cite[(5.4.10)]{Kac} and the locality condition
\begin{equation}
e^{\alpha}c_\alpha e^{\beta}c_\beta = (-1)^{p(\alpha)p(\beta)+
\langle\alpha,\beta\rangle} e^{\beta}c_\beta e^{\alpha}c_\alpha
\end{equation}
Together with free bosonic fields (for all $h\in\mathfrak{h}$)
\begin{equation}
Y((h\otimes t^{-1})\otimes 1, z) = h(z) = \sum_{n\in Z} h_nz^{-n-1}
\end{equation}
they generate a vertex operator algebra with the space of state $V_L$
and the vacuum vector $|0\rangle = 1\otimes 1$ \cite[prop.~5.4]{Kac}.

\smallskip Putting $c_\alpha\left(s\otimes e^\beta\right) =
c_{\alpha,\beta}^{} s\otimes e^\beta$ for $s\in \Sym\left(\mathfrak{h}^{<0}\right)$ and $\beta\in\Lambda$, we get the following equations for numbers $c_{\alpha,\beta}\in\mathbb{C}$:
\begin{align}
 c_{\alpha,0}^{} &=c_{0,\beta}^{}=1\\
 c_{\alpha,\beta}^{} &=(-1)^{p(\alpha)p(\beta)+
\langle\alpha,\beta\rangle} c_{\beta,\alpha}^{}\\
 c_{\beta,\gamma}^{} c_{\beta+\gamma,\alpha}^{}
 &=c_{\gamma,\alpha+\beta}^{} c_{\beta,\alpha}^{}
\end{align}
It defines a cohomology class in $H^2(\Lambda,\mathbb{C}^\times)$, asserting the existence of a unique (up to isomorphism) vertex superalgebra structure on $V_\Lambda$ \cite[sect.~4.4.3]{FBZ}.

\smallskip Moreover, it is a rational conformal superalgebra with central charge $c=\rank(\Lambda)$.

\section{Lattice theta functions}

It can be shown (cf.~\cite[sect.~4.5.5]{FBZ}, \cite[p.~300]{Zhu}) that characters of lattice vertex algebras $V_\Lambda$ of even lattices $\Lambda$ are lattice theta functions (divided by
$\eta(\tau)^{\rank\Lambda}$ where $\eta(\tau)$ is the 
Dedekind $\eta$-function).

\smallskip Let $\Lambda$ be an integral lattice of rank $n$. We define the \emph{theta series} of $\Lambda$ as
\begin{equation}
\Theta_\Lambda(q) = \sum_{v\in\Lambda} q^{\langle v,v\rangle/2} = 
1 + \sum_{k>0} N_{2k}(\Lambda)q^k
\end{equation}
where $N_k(\Lambda) = \#\left\{v\in\Lambda\, |\, \langle v,v
\rangle = k\right\}$. If $k>0$ is the minimal integer such that 
$N_k(\Lambda)>0$ then $N_k(\Lambda)$ is called the \emph{kissing number} of $\Lambda$.

\smallskip The lattice theta-series satisfies the functional equation
\begin{equation}
\Theta_{\Lambda^*}(e^{-2\pi t}) = \sqrt{\disc(\Lambda)}\, 
t^{-n/2} \Theta_\Lambda(e^{-2\pi/t})
\end{equation}
where $\Lambda^*$ is the dual lattice and $\disc(\Lambda)
=\#(\Lambda^*/\Lambda)$ is the discriminant of $\Lambda$.

\smallskip Consider $\theta_\Lambda(\tau)=\Theta_\Lambda
(e^{2\pi i\tau})$ as a function on the upper half-plane 
$\mathbb{H}_+$. If $\Lambda=\Lambda^*$ is self-dual (integral and unimodular) then we get functional equations:
\begin{equation}
\theta_\Lambda(\tau+2)=\theta_\Lambda(\tau),\ \theta_\Lambda(-1/\tau)=(\tau/i)^{n/2}\theta_\Lambda(\tau)\, .
\end{equation}
Recall that the full modular group $\mathrm{PSL}_2(\mathbb{Z})$ is generated by matrices $S=\begin{pmatrix}0 &-1\\ 1 &0\end{pmatrix}$ and $T=\begin{pmatrix}1 &1\\ 0 &1\end{pmatrix}$. Then, for any $\gamma\in\Gamma_+=\langle S,T^2\rangle$, we have
\begin{equation}
\theta_\Lambda(\gamma(\tau))=\varepsilon_{c,d}^n (c\tau+d)^{n/2} \theta_\Lambda(\tau)
\end{equation}
So, $\theta_\Lambda(\tau$) becomes a modular form of weight $n/2$ for 
$\Gamma_+$ with multiplier system $\varepsilon_{c,d}\in\mathbb{C}^*$ 
satisfying $\varepsilon_{c,d}^8=1$ (\emph{cf}.~\cite[sect.~2]{Elk}).

\smallskip Notice that the Dedekind $\eta$-function
\begin{equation}
\eta(\tau) = q^{\frac{1}{24}}\prod_{n=1}^\infty \left(1-q^n\right)
\end{equation}
satisfies functional equations
\begin{equation}
\eta(\tau+1)=e^{\pi i/12}\eta(\tau),\ \eta(-1/\tau)=
(\tau/i)^{1/2}\eta(\tau)\, .
\end{equation}
Thus, $\eta(\tau)$ is a modular form of weight $1/2$ and level 1 with a multiplier system of order 24.

\smallskip The general case of lattices with arbitrary discriminants, related to modular forms for congruence subgroups, needs additional investigations. However, when $\Lambda$ is an even lattice, the modularity follows from Zhu's results on characters of rational vertex algebras \cite{Zhu} (see sect.~6).

\section{Vertex algebras for number fields}

Let $K=\mathbb{Q}[\xi]$ be a Galois number field over $\mathbb{Q}$ of degree $n=r_1+2r_2$ where $r_1$ is the number of real embeddings and $2r_2$ the number of complex embeddings of $K$ in $\mathbb{C}$. In this case of Galois extension over $\mathbb{Q}$, K is either totally real ($\xi\in\mathbb{R}$) with $r_2=0$ or totally imaginary 
($\xi\in\mathbb{C}\backslash\mathbb{R}$) with $r_1=0$.

\smallskip The important thing here is to use the nondegenerate symmetric bilinear form 
\begin{equation}
\mathrm{tr}:K\times K\rightarrow\mathbb{Q},\ \langle x,y\rangle\mapsto tr_{K/\mathbb{Q}}^{}(xy),
\end{equation}
called the \emph{trace form}. When $K$ is totally real, it is positive-definite. In the totally imaginary case, it is negative-definite and we should take rather the anti-trace form $\langle x,y\rangle\mapsto -tr_{K/\mathbb{Q}}^{}(xy)$.

\smallskip Let $\mathcal{O}_K$ be the \emph{ring of integers} in $K/\mathbb{Q}$ corresponding to a lattice $\Lambda$, constructed from real and complex embeddings of $K$ (see \cite[ch.~1, \S~1]{PS}). Recall that $\mathcal{O}_K$ is the unique maximal order in $K$. In our number-theoretical setting, there is a finite number of orders preserving $\Lambda$, equal to the \emph{class number} $h(K)$ of $K$. The class number measures the deviation of 
$\mathcal{O}_K$ from being a principal ideal domain.

\smallskip Our bilinear form $\langle\cdot,\cdot\rangle$ restricts 
to an integral form on $\mathcal{O}_K$. Now, choosing a basis 
$\{\omega_1^{},\dots,\omega_n^{}\}$ of $\mathcal{O}_K$, we can extend $\langle\cdot,\cdot\rangle$ to a positive-definite integral form on 
$\Lambda$.

\smallskip So, we can construct a vertex superalgebra 
$\mathcal{V}_K$, associated to any Galois number field 
$K/\mathbb{Q}$. It opens the whole area of investigations in order to express the class field theory, aritmetical reciprocity laws etc. in terms of modularity theorems for lattice vertex superalgebras.

\smallskip Consider a simple example of a real quadratic field 
$K=\mathbb{Q}\left(\sqrt{d}\right)$ where $d$ is a square-free positive integer. Let
\begin{equation}
\omega=\begin{cases}(1+\sqrt{d})/2 &d\equiv 1\pmod 4 \\ 
\sqrt{d} &d\equiv 2,3\pmod 4\end{cases}
\end{equation}
Then $\{1,\omega\}$ is a basis of $\mathcal{O}_K$ 
\cite[ch.~1, ex.~2]{PS}. An arbitrary order in $K$ is of the form 
$\mathcal{O}_f=\mathbb{Z}[f\omega]$ with a positive integer $f$ called the \emph{conductor} of $\mathcal{O}_f$. If $\alpha=a+b\sqrt{d}$ then the trace $tr_{K/\mathbb{Q}}^{}(\alpha) = 2a$.

\smallskip As we have seen in section 3, the vertex superalgebra structure on $V_\Lambda$ is determined by a cocycle $c_{\alpha,
\beta}\in H^2(\Lambda,\mathbb{C}^\times)$. Kac indicates an explicit form of this cocycle with values in $\mathbb{Z}/2\mathbb{Z}$ 
\cite[sect~5.5]{Kac}.

\smallskip In order to be fully explicit, consider, for instance, 
$K=\mathbb{Q}[\sqrt{2}]$, $\mathcal{O}_K=\mathbb{Z}[\sqrt{2}]$ and 
the corresponding even integral lattice 
$\Lambda=\mathbb{Z}\oplus \mathbb{Z}\sqrt{2}\in\mathbb{R}^2$ with basis $\{\alpha=(1,0),\, \beta^{}=(0,\sqrt{2})\}$. Our bilinear form, associated to the trace form, gives
\begin{align}
\langle\alpha,\alpha\rangle &= tr_{K/\mathbb{Q}}^{}(1)=2\\
\langle\alpha,\beta\rangle =\langle\beta,\alpha\rangle 
 &=tr_{K/\mathbb{Q}}^{}(\sqrt{2}) = 0\\
\langle\beta,\beta\rangle &= tr_{K/\mathbb{Q}}^{}(2)=4
\end{align}
So, the matrix of the trace form is diagonal and \emph{discriminant} $d(K)=2\cdot 4=8$ as expected for this quadratic field. It is also well-known that $h(K)=1$.

\smallskip Thus, the \emph{$2$-cocycle of Frenkel-Kac}, defining the unique (up to isomorphism) lattice vertex algebra $\mathcal{V}_{\mathbb{Z}[\sqrt{2}]}$ (with integral trace form), is given by the following formulas:
\begin{align}
c_{\alpha,\alpha}^{} &= (-1)^{\left((\alpha,\alpha)+
(\alpha,\alpha)^2\right)/2} = (-1)^{(2+2^2)/2} = -1\\ 
c_{\alpha,\beta}^{} &= (-1)^{(\alpha,\beta)+(\alpha,\alpha)
(\beta,\beta)} = 1 = c_{\beta,\alpha}^{}\\
c_{\beta,\beta}^{} &=(-1)^{\left((\beta,\beta)+
(\beta,\beta)^2\right)/2} = (-1)^{(4+4^2)/2} =1\, .
\end{align}
This vertex algebra has $d(K)=8$ inequivalent simple modules.
Indeed, in the case of even integral lattices, Dong \cite{Dong} has shown that inequivalent irreducible modules of 
$\mathcal{V}_\Lambda$ are parameterized by $\Lambda^*/\Lambda$.

\section{Zhu's modularity results for rational CFT}

Let $\mathcal{V}_c$ be a rational conformal vertex algebra of central charge $c$. For any simple $\mathcal{V}$-module $M$ one can define its character by the formula
\begin{equation}
\ch M^{} = \Tr_M^{} q^{L_0^M-c/24}=\sum_\alpha \dim M_\alpha
q^{\alpha - c/24}\, ,
\end{equation}
where $M_\alpha$ is the subspace of M on which the Virasoro operator $L_0^M$ acts by multiplication by $\alpha$.

\smallskip Let $C_2(\mathcal{V}_c)$ be the subspace generated by elements of the form $A_{(-2)}\cdot B$ for all $A,B\in \mathcal{V}_c$. Then $\mathcal{V}_c$ is said to satisfy \emph{Zhu's finiteness condition} if $\dim \mathcal{V}_c/C_2(\mathcal{V}_c)<\infty$ and any vector can be written as $L_{n_1^{}}\cdots L_{n_k^{}}A$, $n_i<0$, 
where $L_nA=0$ for all $n>0$.

\begin{thm}(\cite[thm.~5.3.2]{Zhu})
Let $\mathcal{V}_c$ be a rational conformal vertex algebra of central charge $c$ satisfying Zhu's finiteness condition. Let $\{M_1,\dots,
M_m\}$ be the complete list of irreducible $\mathcal{V}_c$-modules. Consider \emph{$n$-point correlation functions} 
\begin{equation}
S_{M_i}\left((a_1,z_1),\dots,(a_n,z_n),\tau\right)
\end{equation}
as meromorphic continuations ($q=e^{2\pi i \tau}$) of limits
\begin{equation}
\Tr_{M_i} Y(e^{2\pi z_1L_0}a_1,e^{2\pi z_1L_0})\cdots
Y(e^{2\pi z_1L_0}a_n,e^{2\pi z_nL_0})q^{L_0- c/24}
\end{equation}
If $a_1,\dots,a_n$ are highest weight vectors with weights $w_1,\dots,
w_n$ for the Virasoro algebra, then for every $\begin{pmatrix}
a &b\\ f &d\end{pmatrix}\in\mathrm{SL}_2(\mathbb{Z})$, we have
\begin{align}
S_{M_i}\left(\left(a_1,\frac{z_1}{f\tau+d}\right),\dots,
\left(a_n,\frac{z_n}{f\tau+d}\right),\tau\right)=\\
(f\tau+d)^{\sum w_k} \sum_{j=1}^m S(\alpha,i,j)
S_{M_j}\left((a_1,z_1),\dots,(a_n,z_n),\tau\right)
\end{align}
where $S_\alpha(i,j)$ are constants depending only on $\alpha,i,j$.
In particular, if $\mathcal{V}_c$ has a unique simple module $M$, and $a$ is a highest weight vector of weight $w$ for the Virasoro algebra, then $S_M(a,\tau)$ is a modular form of weight $w$ with a certain multiplier system.
\end{thm}

\smallskip Here we see the appearance of the \emph{modular $S$-matrix} 
$S(\alpha,i,j)$ for any $\alpha\in\mathrm{SL}_2(\mathbb{Z})$. The functional equation for correlation functions corresponds to $\alpha=\begin{pmatrix} 0 &-1\\ 1 &0\end{pmatrix}$.

\begin{thm}(\cite[thm.~5.3.3]{Zhu})
In the conditions of the previous theorem, $\ch M_i$ converges to holomorphic functions on $\mathcal{H}_+$ and the space, spanned by 
$\ch M_i$, $1\leqslant i\leqslant n$, is invariant under the action
of $\mathrm{SL}_2(\mathbb{Z})$.
\end{thm}

\section{Minimal models and $S$-matrix of the critical Ising model}

It is known that $\mathrm{Vir}_c$ is reducible as the module over the Virasoro algebra if and only if
\begin{equation}
c=c(p,r)=1-6(p-r)^2/pr,\ p,r>1,\ (p,r)=1
\end{equation}
In this case, the irreducible quotient $L_{c(p,r)}$ of 
$\mathrm{Vir}_{c(p,r)}$ is a rational vertex algebra, called 
\emph{minimal model} of Belavin-Polyakov-Zamolodchikov \cite{BPZ, dFMS,Wang}.

\smallskip In order to be explicit, consider the case $(p,r)=(4,3)$ with $c=1/2$. It is known as $d=2$ \emph{critical Ising model}. There are 3 simple modules $M_0$, $M_{1/16}$ and $M_{1/2}$ with conformal dimensions $0$, $1/16$ and $1/2$ correspondings to primary fields $1$ (identity), $\sigma$ (spin) and $\epsilon$ (energy).

\smallskip The characters are given by the following formulas 
($q=e^{2\pi i \tau})$:
\begin{align}
\ch M_0 &= \frac{\sqrt{\theta_3(0|q)}+\sqrt{\theta_4(0|q)}}
{2\sqrt{\eta(\tau)}}\\
\ch M_{\frac{1}{16}} &= \frac{\sqrt{\theta_3(0|q)}-\sqrt{\theta_4(0|q)}}{2\sqrt{\eta(\tau)}}\\
\ch M_{\frac{1}{2}} &= \frac{\sqrt{\theta_2(0|q)}}{\sqrt{2\eta(\tau)}}\, .
\end{align}
Here we used 3 remarquable theta functions (where $q=e^{\pi i \tau}$):
\begin{equation}
 \theta_2(0|q)=\frac{2\eta^2(2\tau)}{\eta(\tau)},\
 \theta_3(0|q)=\frac{\eta^2((\tau+1)/2)}{\eta(\tau+1)},\
 \theta_4(0|q)=\frac{\eta^2(\tau/2)}{\eta(\tau)}\, .
\end{equation}
The functional equation for characters:
\begin{equation}
\ch M_i (-1/\tau) = \sum_j S(i,j)\ch M_j(\tau)
\end{equation}
is expressed in terms of the modular $S$-matrix: 
\begin{equation}
S(i,j)=\frac{1}{2}\begin{pmatrix} 1 &1 &\sqrt{2}\\
1 &1 &-\sqrt{2}\\ \sqrt{2} &-\sqrt{2} &0\end{pmatrix} .
\end{equation}
corresponding to $\alpha=\begin{pmatrix} 0 &-1\\ 1 &0\end{pmatrix}$ in the Zhu's theorem.

\smallskip Finally, notice that the \emph{partition function}
\begin{equation}
Z(q) = \sum_i |\ch M_i|^2
\end{equation}
is modular invariant.

\section{Hecke eigenforms and Wiles modularity theorem}

The $m$th Hecke operator $T_m$ acts on lattice functions by taking sums over all sublattices of index $m$. If $f(z)=\sum a_n^{}q^n$ a weakly holomorphic (with possible poles at cusps) modular form of weight $k$ then

\begin{align}
T_m f(z) &= n^{k-1}\sum_{a,d>0,\, ad=n} \frac{1}{d^k}
\sum_{b\,\mathrm{mod}\, d} f\left(\frac{az+b}{d}\right)\\
&= \sum_{ad=n,\, \mu>0} a^{k-1}c_{d\mu}^{}q^{a\mu}
\end{align}
Denote $M_k$ the space of such weakly holomorphic modular forms and $S_k$ the space of cusp forms with $a_0^{}=0$. Hecke operators transform cusp forms into cusp forms and, for a prime $m=p$, we get $T_p(f)=\sum b_nq^n$ with
\begin{equation}
b_n=\begin{cases}a_{pn}^{} &\mathrm{if}\ p\nmid n\\ 
a_{pn}^{}+p^{k-1}a_{n/p} &\mathrm{if}\ p\mid n \end{cases}
\end{equation}
We can present $S_k$ as the direct sum $S_k^{\mathrm{old}}\oplus
S_k^{\mathrm{new}}$ of subspaces of old and new cusp forms, orthogonal to each other with respect to the Petersson inner product. Old forms arise from modular forms of lower levels and have $a_1=0$. New forms with the same Diriclet character $\chi\pmod N$ are proportional and could be normalized by putting $a_1=1$. This is known as \emph{multiplicity 1 theorem}.

\smallskip Denote $\Gamma_0(N)$ the subgroup of matrices $\begin{pmatrix}
a &b\\ c &d\end{pmatrix}\in\mathrm{SL}_2(\mathbb{Z})$ with $c\equiv
0\pmod{N}$ and consider the modular curve $X_0(N)=\Gamma_0(N)\backslash
\mathbb{Z}_+$.

\begin{thm}(Wiles et al.) Let $E/\mathbb{Q}$ be a semi-stable elliptic curve with conductor $N$ and $L$-function
\begin{equation}
L(E,s) = \sum_{n=1}^{\infty} \frac{a_n}{n^s}\, .
\end{equation}
Then there exists a new cusp Hecke eigenform $f(\tau)=\sum{b_nq^n}$ of weight 2 and level $N$ with $b_n=a_n$ (for almost all $n$), defining a modular parameterization $X_0(N)\rightarrow E$ (up to an isogeny).
\end{thm}

\section{Hecke eigenfields and Maeda's conjecture}

\smallskip Let $f$ be a new cusp Hecke eigenform of level $N$ then
\begin{equation}
T_m f(\tau) = \lambda_m f(\tau),\ \lambda_m = a_m^{}a_1^{}
\end{equation}
When $f(\tau)$ is normalized, we have $\lambda_m=a_m^{}$.

\smallskip Consider the field $K_f=\mathbb{Q}[\lambda_m^{},m\geqslant 1]$, generated by Hecke eigenvalues, that will be called \emph{Hecke eigenfield}, associated to $f$. It is an algebraic extension of 
$\mathbb{Q}$.

\begin{prop} Let $N=1$. The Hecke eigenfield $K_f$ is totally real or CM-field (when $a_1^{}\in\mathbb{C}\backslash\mathbb{R}$).
\end{prop}

\pn \textsl{Proof.} It follows from the fact that Hecke operators are self-adjoint with respect to the Petersson inner product. $\square$

\smallskip Shimura has attached to $f(\tau)$ of weight $2$ an abelian variety $\mathrm{Sh}_f$ of dimension $[K_f:\mathbb{Q}]$. In the Wiles case of elliptic curve $E/Q$, $K_f=\mathbb{Q}$ and $Sh_f=E$ (up to an isogeny). Hecke eigenvalues correspond to the monodromy of the Galois action on 
$f(\tau)$ (as a horizontal section of an appropriate line bundle with a flat connection) around cusps. This is one of basic keypoints of the 
Beilinson-Drinfeld geometric Langlands duality.

\smallskip The following \emph{Maeda's conjecture} \cite[sect.~3.6.1]{RS} is important in order to to study the Langlands duality in terms of Hecke eigenfields.

\begin{conj}
There is only one $Gal(\overline{\mathbb{Q}}/\mathbb{Q})$-orbit of normalized eigenforms of level 1.
\end{conj}

\sn It is numerically proven for all $k\leqslant 4096$ \cite{RS}.

\section{Critical WZW and Beilinson-Drinfeld geometric correspondence}

There are several books written on this topic \cite{BD, Fr}, so we briefly underline main ideas. It may be considered as the geometric $S$-duality for WZW-models for affine Kac-Moody algebras $\widehat{G}$ at critical level $c=h^{\vee}$ (=dual Coxeter number of $G$).

\smallskip Let $G$ be a reductive Lie group and $^LG$ its Langlands dual (with dual root system). Denote $\mathrm{Bun}_G^{}(X)$ the moduli space of principal $G$-bundles on a curve $X/\mathbb{C}$, $g(X)\geqslant 2$.

\smallskip According to Beilinson and Drinfeld, there is a one-to-one correspondence 
\begin{equation}
\textrm{Hecke eigensheaves on Bun$_G^{}(X)$} \leftrightarrow 
\textrm{$^LG$-local systems on $X$}
\end{equation}
Moreover, these Hecke eigensheaves are $D$-modules and their ``eigenvalues'' are exactly the corresponding $^LG$-local systems. Via the \emph{Riemann-Hilbert correspondence}, these local systems are representations of the fundamental group $\pi_1(X)$ playing the role of Galois representations.

\smallskip Notice that geometric versions of Hecke operators are crucial for the geometric Langlands theory. 

\section{Witten's gravitational moonshine}

Let $\mathcal{A}$ be a $3d$-spacetime asymptotic at infinity to AdS$_3$. Witten has suggested that the pure quantum gravity on $\mathcal{A}$ with maximally negative cosmological constant corresponds to an extremal $c=24$ two-dimensional CFT. This remarquable CFT was constructed by Frenkel-Lepowsky-Meur\-man \cite{FLM} as the so-called \emph{Moonshine Module} $V^\natural$ of the \emph{Monster vertex algebra} $\mathbb{M}$. $\mathbb{M}$ is a rational vertex algebra and its unique irreducible  representation $V^\natural$ is just $\mathbb{M}$ as module over itself. In this case there are 196883 operators, associated to Virasoro primary fields of $V^\natural$, creating BTZ black holes on $\mathcal{A}$ \cite{Wit}.

\smallskip The partition function of $c=24$ pure quantum gravity on AdS$_3$ is simply
\begin{equation}
Z_1(q)=J(q)=j(q)-744
\end{equation}
where $j(q)$ is the absolute $j$-invariant. Partition functions $Z_k(q)$ of hypothetical $c=24k$ quantum gravities can be obtained using Hecke operators $T'_m= mT_m$ (or Faber polynomials $\Phi_m(X)$ such that $\Phi_m(j)=j_m(q)=T'_m(J)$ \cite{Gue}). Witten shows \cite[(3.13)]{Wit}
that $Z_k(q)$ should be of the form
\begin{equation}
Z_k(q)=\sum_{m=0}^k a_{-m} T'_mJ(q)=\sum_{m=0}^k a_{-m} j_m(q)
\end{equation}
where $a_{-m}$, $0\leqslant m\leqslant k$, come from the formula
\begin{equation}
q^{-k}\prod_{n=2}^\infty (1-q^n)^{-1}=\sum_{m=-k}^\infty
a_m q^m\, .
\end{equation}
In this way we obtain
\begin{equation}
Z_2(q)=(T'_2+T'_0)J(q)
\end{equation}
and so on. This construction is consistent with the Bekenstein-Hawking entropy for black holes. However, it is not yet clear to what extremal 
CFT correspond partitions functions $Z_k(q)$ for $k>1$.

\smallskip It looks relatively simple at the level of partition functions but this is misleading. Zhu \cite{Zhu} shows that $V^\natural$ can be decomposed as the direct sum of tensor products of highest weight modules of 48 Virasoro algebras with central charge $c=1/2$. Actually, it also proves the rationality of $\mathbb{M}$.

\section{Perspectives}

We have indicated just the beginning of the story. First of all, number-theoretical questions can be treated as a particular case of the study of lattice vertex superalgebras for number fields. It demonstrates an amazing unification of the arithmetic with lattice CFT theories.

\smallskip On the one hand, arithmetic generalizations would include Drinfeld associators, motivic Galois groups and Grothendieck-Teichm\"uller groupoids. In positive characteristic, an upcoming article \cite{Po2} will treat vertex $t$-algebras, generalizing Drinfeld modules and Anderson's 
$t$-motives.

\smallskip On the other hand, physical generalisations would include the 
$S$-duali\-ty between electric and magnetic branes of the unifying
$M$-theory. As Witten have already noticed, Hecke operators are related 
to 't Hooft operators and Wilson loops.

\smallskip The holographic preimage of the Monstrous Moonshine, giving a spacetime with BTZ black holes, is inspiring and requires additional investiations.

\mn {\scriptsize \textbf{Acknowledgments.} I would like to thank my colleagues Vadim Schechtman and Joseph Tapia as well as my old friend 
Pavel Guerzhoy for inspiring discussions.}

\nocite{*}
\bibliographystyle{alpha}
\bibliography{gravshine2}

\begin{thebibliography}{dFMS97}

\bibitem[Ash20]{Ash}
M.~Ashrafi.
\newblock Three dimensional gravity and the generalized hecke operators.
\newblock {\em arXiv: 2004.04424v3}, pages 1--20, 2020.

\bibitem[BD04]{BD}
A.~Beilinson and V.~Drinfeld.
\newblock {\em Chiral Algebras}.
\newblock AMS, 2004.

\bibitem[BPZ84]{BPZ}
A.~Belavin, A.~Polyakov, and A.~Zamolodchikov.
\newblock Infinite conformal symmetries in two-dimensuional quantum field
  theory.
\newblock {\em Nucl. Phys B}, pages 333--380, 1984.

\bibitem[dFMS97]{dFMS}
P.~di~Francesco, P.~Mathieu, and D.~Senechal.
\newblock {\em Conformal Field Theory}.
\newblock Springer, 1997.

\bibitem[Don93]{Dong}
C.~Dong.
\newblock Vertex algebras associated with even lattices.
\newblock {\em J. Algebra}, pages 245--265, 1993.

\bibitem[Elk09]{Elk}
N.~Elkies.
\newblock Theta functions and weighted theta functions of {E}uclidean lattices,
  with some applications.
\newblock unpublished, 2009.

\bibitem[FBZ01]{FBZ}
E.~Frenkel and D.~Ben-Zvi.
\newblock {\em Vertex algebras and algebraic curves}.
\newblock AMS, 2001.

\bibitem[FLM88]{FLM}
I.~Frenkel, J.~Lepowsky, and A.~Meurman.
\newblock {\em Vertex operator algebras and the {M}onster}.
\newblock Academic Press, 1988.

\bibitem[Fre07]{Fr}
E.~Frenkel.
\newblock {\em Langlands corresondence for loop groups}.
\newblock Cambridge University Press, 2007.

\bibitem[Gue08]{Gue}
P.~Guerzhoy.
\newblock On irreducibility of certain {F}aber plynomials.
\newblock {\em Ramanujan J.}, 1:53--57, 2008.

\bibitem[GW07]{GW}
S.~Gukov and E.~Witten.
\newblock Gauge theory, ramification and geometric {L}anglands program.
\newblock {\em arXiv: hep-th/0612073v2}, pages 1--159, 2007.

\bibitem[Kac98]{Kac}
V.~Kac.
\newblock {\em Vertex algebras for beginners (2nd edition)}.
\newblock AMS, 1998.

\bibitem[Koc92]{PS}
H.~Koch.
\newblock Encyclopaedia of {M}athematical {S}ciences.
\newblock In I.R.~Shafarevich A.N.~Parshin, editor, {\em Number theory II}.
  Springer, 1992.

\bibitem[KW01]{KW}
A.~Kapustin and E.~Witten.
\newblock Electric-magnetic duality and the geometric {L}anglands program.
\newblock {\em Communications in number theory and physics}, 1:1--16, 2001.

\bibitem[Pot99]{Po1}
I.~Potemine.
\newblock Drinfeld-{A}nderson motives and multicomponent {KP}-hierarchy.
\newblock In {\em Contemporary Mathematics 224}, pages 213--227. AMS, 1999.

\bibitem[Pot01]{Po2}
I.~Potemine.
\newblock Quantum $\textrm{L}$anglands duality and mirror symmetry.
\newblock {\em arXiv: math/0111260}, pages 1--16, 2001.

\bibitem[Pot21]{Po3}
I.~Potemine.
\newblock Vertex $t$-algebras in positive characteristic.
\newblock in preparation, 2021.

\bibitem[RS17]{RS}
K.~Ribet and W.~Stein.
\newblock Lectures on modular forms and {H}ecke operators.
\newblock unpublished, 2017.

\bibitem[Wan93]{Wang}
W.~Wang.
\newblock Rationality of {V}irasoro vertex operator algebras.
\newblock {\em IMRN}, pages 197--211, 1993.

\bibitem[Wit07]{Wit}
E.~Witten.
\newblock Three-dimensional gravity reconsidered.
\newblock {\em arXiv: 0706.3359}, pages 1--82, 2007.

\bibitem[Zhu96]{Zhu}
Y.~Zhu.
\newblock Modular invariance of characters of vertex operator algebras.
\newblock {\em J. AMS}, pages 237--302, 1996.

\end{thebibliography}

\bigskip
\begin{flushright}
Igor Potemine\\
Laboratoire Emile Picard\\
Universit\'e Paul Sabatier\\
118, route de Narbonne\\
31062 Toulouse (France)
\end{flushright}

\begin{flushright}
e-mail : igor.potemine@math.univ-toulouse.fr
\end{flushright}

\end{document}